\documentstyle[pre,aps]{revtex}

\begin{document}

\draft
\title{STATISTICAL MECHANICS OF STRUCTURAL FLUCTUATIONS}
\author{V. I. Yukalov} 
\address{Department of Mathematics and Statistics \\
Queen's University, Kingston, Ontario K7L 3N6 \\
Canada}
\maketitle

\begin{abstract}

The theory of mesoscopic fluctuations is applied to inhomogeneous solids 
consisting of chaotically distributed regions with different crystalline 
structure. This approach makes it possible to describe statistical properties 
of such mixture by constructing a renormalized Hamiltonian. The relative 
volumes occupied by each of the coexisting structures define the corresponding
geometric probabilities. In the case of a frozen heterophase system these 
probabilities should be given a priori. And in the case of a thermal 
heterophase mixture the structural probabilities are to be defined 
self--consistently by minimizing a thermodynamical potential. This permits 
to find the temperature behavior of the probabilities which is especially 
important near the points of structural phase transitions. The presense of 
these structural fluctuations yields a softening of a crystal and a decrease 
of the effective Debye temperature. These effects can be directly seen by 
nuclear gamma resonance since the occurrence of structural fluctuations 
is accompanied by a noticeable sagging of the M\"ossbauer factor at the point 
of structural phase transition. The structural fluctuations also lead to the 
attenuation of sound and increase of isothermic compressibility.

\end{abstract}

\section{Introduction}   

There are many examples of matter consisting of regions, chaotically 
distributed in space, with different structural properties. For instance, 
such are some polymorphic materials. Another example is a crystal subject 
to strong mechanical stress after which the cracks and branches of dislocation
are formed in it. These defects have a tendency to group inside compact 
regions. The latter, from the point of view of statistical physics, can 
be treated as nuclei of the amorphised phase inside a crystalline matrix 
[1]. A similar picture develops in crystals under the action of irradiation 
by fast neutrons when the pores and regions of disorder arise. Under strong 
irradiation cracks also appear. These defects form groups and clusters 
randomly distributed in space, e.g. as is shown in Figs. 1 and 2. For a 
statistical description of an irradiated crystal the defected regions can 
be treated as embryos of disordered, usually rarefied, phase inside an 
ordered, more dense, crystalline structure [2,3]. The relative volume 
occupied by the disordered phase can be measured, with a good accuracy, 
by investigating the nuclear gamma resonance spectra and the behavior of 
the M\"ossbauer factor [4,5].

In the considered examples the germs of a disordered structure are randomly 
distributed in space inside an ordered structure. This is why these germs can 
be called the spatial structural fluctuations. With respect to time, they are 
frozen, which means that their average lifetime, $ \; \tau_f \; $, is much 
longer than the characteristic time of an experiment, or the observation time,
$\;\tau_{obs}\;$, that is: $\;\tau_f\gg\tau_{obs}\;$. In the opposite case, 
when $\;\tau_f \ll \tau_{obs} \;$, we have thermal structural fluctuations. 
The example of the latter are even more numerous than those of the frozen 
structural fluctuations.

Water, the most widespread matter on Earth, gives us a cogent example of a 
system consisting of at least two coexisting structures, which is supported 
by numerous experiments studying its thermodynamic and dielectric properties 
and analysing the Raman spectroscopy data [6 - 8]. A very convincing is the 
molecular - dynamic investigation [9] showing that fluctuations in water can 
be decomposed into two components: a fast component 
($\; 10^{-14}-10^{-13}s\;$) associated with the libration motion in one 
of the water inherent structures and a slow component 
($\; 10^{-12}-10^{-11}s\;$) associated with the water structure changes. 
These thermal structural fluctuations in water are related to large local 
energy fluctuations involving about 10--100 molecules.

Thermal structural fluctuations often appear in the vicinity of structural 
phase transitions. In principle, these fluctuations are possible near both 
types of structural transitions, near displacive as well as near 
order--disorder type. The distinction between the two types of structural 
phase transitions can be traced, for example, by their different isotope 
effects [10], although in the presense of strong fluctuations this distinction
becomes less pronounced [11 - 13]. The existence of structural fluctuations 
around the point of a phase transition reveals itself in the so--called 
pretransitional phenomena [14,15].

The freezing - melting phase transition is also a kind of structural 
transitions. Below the melting point, the fluctuational lattice dissociation 
can arise breaking the local crystalline symmetry, while above the freezing 
point the structural fluctuations are represented by quasicrystalline clusters
inside a fluid phase with no crystalline structure. Only taking into account 
the existence of these structural fluctuations it is possible to develop a 
realistic theory of melting and crystallization [16,17]. Such fluctuations 
should be especially important during the melting of small clusters [18] and 
solid--fluid phase transition of quantum systems [19].

Quantum crystals exhibit near phase transition points the well observed 
fluctuational coexistence of competing structures. This takes place, for 
example, in solid $ \; ^4He \; $ along the line of the structural transition 
between h.c.p. and b.c.c. phases and also in the solid mixture of $\;^4He\;$
with $ \; ^3He \; $ along the stratification line [20]. These local 
fluctuational effects involve usually about 100 particles.

Fluctuations of one phase inside another can be called [21] the heterophase 
fluctuations. The structural fluctuations are just a sort of the latter. An 
extensive description of a great number of systems exhibiting these 
fluctuations has been done in the recent review [22]. Therefore, we now shall 
limit ourselves by the examples considered above, mentioning only that the 
structural fluctuations can play a crucial role in high - temperature 
superconductors [23]. 

The heterophase fluctuations are mesoscopic since the characteristic size 
of such a fluctuation, $ \; l_f \; $, satisfies the inequality
$$ a \ll l_f \ll L , $$
in which $ \; a \; $ is the average distance between particles of the system 
and $ \; L \; $ is the linear size of the latter. This drastically 
distinguishes the mesoscopic structural fluctuations from the microscopic 
fluctuations of particles inside the same structure. Thus, the oscillations 
of particles in a crystal in the vicinity of the corresponding lattice sites 
are the most known microscopic fluctuations, the root - mean - square 
deviation of a particle from its lattice position being $\;r_0\ll a\;$.

Another characteristic feature of the heterophase fluctuations is that 
the typical energy of each of them is much larger than the average 
single - particle energy [24]. This seems to be quite understandable 
since a heterophase fluctuation, being mesoscopic by its nature, involves 
many particles whose number is $ \; N_f \gg 1 \; $.

The system with structural fluctuations are  difficult to describe. This is 
because such systems are nonuniform with the nonuniformity occurring on 
mesoscopic scale [22]. Each structural nucleus can have a complicated 
ramified form and can exhibit nontrivial motion [24]. Some thermodynamic 
features of such systems can be seized by simple phenomenological models 
(see discussion in Ref.[22] and also [25,26]). However, in phenomenological 
treatment one usually has to invoke one or several fitting functions designed 
to satisfy particular experiments. Of course, this is not satisfactory from 
the point of view of statistical mechanics. These and other difficulties have 
been discussed and illustrated by a mechanical model in Ref.[27].

Recently, a consistent statistical approach has been developed for treating 
the systems with mesoscopic heterophase fluctuations [22]. In the present 
paper we apply this approach to the systems with structural fluctuations. 
In Sec.II the main steps of constructing a renormalized Hamiltonian for such 
systems are explained, the general equations for the structural probabilities 
and stability conditions are analysed. In Sec.III we concretize the problem 
for the coexisting crystalline structures. The behavior of structure 
probabilities in the vicinity of a temperature of a structural phase 
transition is studied in Sec.IV, using the Debye approximation. A very 
effective tool for examining the properties of structural fluctuations is 
the M\"ossbauer effect. Therefore, in Sec.V we scrutinize the peculiar 
behavior of the M\"ossbauer factor near the temperature of structural phase 
transition and show how the presense of structural fluctuations yields 
characteristic saggings of the M\"ossbauer factor. These fluctuations also 
lead to the attenuation of the sound velocity and to the enhancement of the 
isothermal compressibility, as is discussed in Sec.VI. R\'esum\'e is given 
in Sec.VII. Everywhere below the system of units is used where 
$\hbar=k_B=1\;$.

\section{Renormalized Hamiltonian}

Consider a system in which two structures can coexist. We enumerate the 
structures by the index $ \; \nu = 1,2 \; $. In the case when both structures 
are crystalline, each of them is characterized by a set 
$$ {\bf A}_\nu = \{ \stackrel{\rightarrow}{a}_{i\nu} | \; i=1,2,\ldots N \}
\qquad (\nu = 1,2) \; $$
of lattice vectors $ \stackrel{\rightarrow}{a}_{i\nu} \; $. As is discused 
above, the distribution of structures in the space is random. Therefore, it 
is necessary to define the procedure of averaging over structure 
configurations. Each configuration can be given by specifying which regions 
$\;{\bf V}_\nu \; (\nu = 1,2)\;$ of the total system volume $\;{\bf V}\;$ 
are occupied by this or that structure, so that
$$ {\bf V} = \cup_\nu {\bf V}_\nu = {\bf V}_1 \cup {\bf V}_2 . $$
Such a specification can be defined [22] by fixing a set 
\begin{equation}
\xi \equiv \{ \xi_\nu  (\stackrel{\rightarrow}{r}) | \; \nu =1,2; \; 
\stackrel{\rightarrow}{r} \in {\bf V} \} 
\end{equation}
of functions
\begin{eqnarray}
\xi_\nu(\stackrel{\rightarrow}{\xi}) \equiv \left \{ \begin{array}{cc}
1 ,  & \stackrel{\rightarrow}{r} \in {\bf V}_\nu \\
0 ,  &  \stackrel{\rightarrow}{r} \not \in {\bf V}_\nu \end{array}
\right.
\end{eqnarray}
that are called the manifold characteristic functions, or the manifold 
indicator functions, or simply, the manifold indicators. In this way, a 
distribution of structures in space, that is a structure configuration, 
is uniquely defined by a covering $\;\{ {\bf V}_\nu |\;\nu =1,2\}\;$ of 
$ \; {\bf V} \; $, or, equivalently, by the indicator set (1) of manifold 
indicators (2). The many of all possible collections of $ \; \xi \; $ form 
the topological space $ \; {\cal T}(\xi ) \; $, In order to define the 
averaging over structure configurations, we need to introduce a functional 
measure on $ \; {\cal T}(\xi ) \; $. This procedure with all mathematical 
details has been thoroughly described in a series of papers [17,28 - 30] 
and expounded in a recent review [22]. Therefore, there is no need to repeat 
it here. But for the logical self - consistency of the present exposition we 
will delineate the main steps of this procedure.

Each fixed structure configuration depicts a nonuniform system which can be 
characterized by the quasiequilibrium Gibbs ensemble with a distribution 
proportional to $ \; e^{-\Gamma (\xi )} \; $, where $ \; \Gamma (\xi ) \; $ 
is a quasi - Hamiltonian defined for a given set (1). Specifying a functional 
measure $ \; \int D\xi \; $ on the topological space of structure 
configurations, we can write the partition function as 
\begin{equation}
Z = Tr\int e^{-\Gamma (\xi )}D\xi ,
\end{equation}
where $ \; Tr \; $ means the trace over all quantum - mechanical degrees of 
freedom, or over the phase space in the classical case. The renormalized 
Hamiltonian $ \; \stackrel{-}{H} \; $ is defined by the relation
$$ \int e^{-\Gamma (\xi )}D\xi = e^{-\stackrel{-}{H}/T} , $$
which yields
\begin{equation}
\stackrel{-}{H} = -T\ln \int e^{-\Gamma (\xi )}D\xi .
\end{equation}
Then the partition function (3) becomes
$$ Z = Tr e^{-\stackrel{-}{H}/T} . $$

For each particular structure a space $ \; {\cal H}_\nu \; $ of microscopic 
states is to be defined, consisting of wave functions enjoying the property 
of the corresponding structural symmetry. The latter, in the case of a 
crystalline structure, is the symmetry of a given crystalline lattice. 
The renormalized Hamiltonian (4) acts on the fiber space
\begin{equation}
{\cal Y} =\otimes_\nu {\cal H}_\nu = {\cal H}_1 \otimes {\cal H}_2 .
\end{equation}
From definition (4) it is clear that the renormalized Hamiltonian should 
depend on the probability weights of coexisting structures. These structure 
probabilities $ \; w_\nu \; $ have the meaning of the geometric probabilities,
that is, each $ \; w_\nu \; $ determines the ratio of the effective volume 
occupied by the phase $ \; \nu \; $ to the total volume of the system. 
According to this definition, the standard probability properties
\begin{equation}
0 \leq w_\nu \leq 1 , \qquad \sum_{\nu} w_\nu = w_1 + w_2 = 1
\end{equation}
hold. In the case of a frozen structure distribution the structure 
probabilities $ \; w_\nu \; $ are to be given a priori. For thermal 
structural fluctuations the values of $ \; w_\nu \; $ are to be found 
from the minimization of a thermodynamic potential
\begin{equation}
f = -\frac{T}{N}\ln Tr e^{-\stackrel{-}{H}/T}
\end{equation}
under condition (6). Taking into account the normalization condition for 
$ \; w_\nu \; $ explicitly, we may write
\begin{equation}
w_1 \equiv w , \qquad w_2 \equiv 1 - w .
\end{equation}
Thence, the extremum of (7) with respect to $ \; w \; $ is given by the 
equation
\begin{equation}
\frac{\partial f}{\partial w} = \frac{1}{N} <\frac{\partial 
\stackrel{-}{H}}{\partial w}> = 0 ,
\end{equation}
where $ \; <\ldots > \; $ implies the statistical average which for an 
operator $ \; \hat A \; $ is written as
$$ <\hat A > \equiv Tr\hat \rho \hat A , \qquad 
\hat \rho \equiv \frac{1}{Z} e^{-\stackrel{-}{H}/T} . $$
Remind that according to (4) the renormalized Hamiltonian 
$ \; \stackrel{-}{H} = \stackrel{-}{H} ( \{ w_\nu \} ) \; $ depends on 
structure probabilities. In addition to Eq.(9) defining $ \; w_\nu \; $, we 
have the inequality
\begin{equation}
\frac{\partial^2 f}{\partial w^2} > 0 \qquad ( 0 \leq w \leq 1 ) 
\end{equation}
showing that the found $ \; w_\nu \; $ provide us with the minimum of (7). 
Eq.(10) is the stability condition with respect to variations of structure 
probabilities. From (7) and (10) we have
\begin{equation}
\frac{1}{N} \left \{ < \frac{\partial^2 \stackrel{-}{H}}{\partial w^2} > - 
\frac{1}{T} < \left ( \frac{\partial \stackrel{-}{H}}{\partial w} 
\right )^2 > \right \} > 0 .
\end{equation}
As far as the second term in (11) is always non - negative, the necessary 
stability condition is
\begin{equation}
\left \{ < \frac{\partial^2 \stackrel{-}{H}}{\partial w^2} > \right \} > 0 .
\end{equation}

To proceed further, we have to concretize the situation. Denote by 
$ \; \stackrel{\rightarrow}{R}_{i\nu} \; $ the position of a particle 
$\;i\;$ in a structure $\;\nu\;$ and by 
$\;\stackrel{\rightarrow}{p}_{i\nu}\;$ the momentum of this particle in the 
same structure; here $ \; i = 1,2,\ldots , N \; $ and $ \; \nu = 1,2 \; $. 
Let $ \; \Phi (\stackrel{\rightarrow}{R}_{ij\nu} ) \; $ be the potential of 
interaction between the particles $ \; i \; $ and $ \; j \; $ for which
\begin{equation}
\stackrel{\rightarrow}{R}_{ij\nu} \equiv \stackrel{\rightarrow}{R}_{i\nu} - 
\stackrel{\rightarrow}{R}_{j\nu} .
\end{equation}
Realizing the procedure described above, after the averaging over structure 
configurations we obtain [22] the renormalized Hamiltonian
\begin{equation}
\stackrel{-}{H} = \oplus_\nu H_\nu = H_1 \oplus H_2 ,
\end{equation}
which is a direct sum of terms
\begin{equation}
H_\nu = w_\nu\sum_{i=1}^{N}\frac{\stackrel{\rightarrow}{p}_{i\nu}^2}{2m} + 
\frac{w_\nu^2}{2}\sum_{i\neq j}^{N}
\Phi (\stackrel{\rightarrow}{R}_{ij\nu} ) ,
\end{equation}
where $ \; m \; $ is a particle mass.

With the renormalized Hamiltonian given by (14) and (15), Eq.(9) yields
\begin{equation}
w = \frac{2\Phi_2 + K_2 - K_1}{2(\Phi_1 + \Phi_2)} ,
\end{equation}
where
\begin{equation}
K_\nu \equiv < \frac{1}{N}\sum_{i=1}^{N}
\frac{\stackrel{\rightarrow}{p}_{i\nu}^2}{2m} >
\end{equation}
is the mean kinetic energy per particle in the structure $ \; \nu \; $ and
\begin{equation}
\Phi_\nu \equiv < \frac{1}{2N}\sum_{i \neq j}^{N}
\Phi ( \stackrel{\rightarrow}{R}_{ij\nu} ) > 
\end{equation}
is the mean potential energy per particle in the same structure. The 
necessary condition (12) gives
\begin{equation}
\Phi_1 + \Phi_2 > 0 .
\end{equation}
Additional stability conditions follows from the requirement that 
$ \; 0 \leq w_\nu \leq 1 \; $, which leads to the inequalities
\begin{equation}
-2\Phi_1 \leq K_1 - K_2 \leq 2\Phi_2 .
\end{equation}
Condition (20) has to be true always since it secures the correct definition 
(6) of the structure probabilities. If (20) does not hold, this means that 
the considered structures cannot coexist, even in a metastable mixed state, 
and the system is to be in a pure state corresponding to the structure that 
provides the minimum of the thermodynamic potential (7). While when conditions
(11),(12) or (19) are not valid but (20) holds, the system with structural 
fluctuations can exist in a metastable state.

The described approach has been applied to different systems (see [22]). It 
has made it possible to construct a consistent theory of melting and 
crystalization [16,17] by considering the coexisting solid and liquid 
structures. It is possible to note that when applying this approach to 
one - and two - dimensional crystals we immediately come to the conclusion 
that in these crystals the infinite long - range order is impossible. Really, 
suppose that the index $ \; \nu = 1 \; $ corresponds to a crystalline 
periodic structure, and $ \; \nu = 2 \; $, to a uniform liquid structure. 
As is known after Peierls [31], the phonon potential energy of one - and 
two - dimensional crystals diverges. In our case this means that 
$ \; \Phi_1 \rightarrow \infty \; $. Then from (16) it follows that 
$ \; w \equiv w_1 \rightarrow 0 \; $, and, respectively, 
$ \; w_2 \rightarrow 1 \; $. This shows that one - and two - dimensional 
periodic structures have zero probability to exist.

The stability condition (19) has a simple physical interpretation permitting 
to understand which structures can, in principle, coexist. All structures can 
be roughly divided into two sorts, rigid and soft, according to the sign of 
$ \; \Phi_\nu \; $. The structure can be called rigid if $\;\Phi_\nu < 0\;$ 
since this implies that the particles are strongly bound. And we can say that 
the structure is soft if $ \; \Phi_\nu > 0 \; $ since the particles forming 
it are weakly bound or unbound. The stability condition (19) tells us that 
two structures can thermally coexist only when at least one of them is soft.
Two rigid structures cannot thermally coexist in a stable system. For example,
liquid has  a soft structure. Therefore, a periodic crystalline structure and 
uniform liquid structure can, in principle, coexist near the solid - fluid 
phase transition point [16,17]. Several examples of possible thermal 
coexistence of different phases have been considered for lattice - gas an 
spin models (see review [22]). Thus, for a lattice - gas model containing 
two phases with different densities it has been shown that such a system is 
unstable when the effective interaction between particles is attractive [2] 
but the system becomes stable if the effective interaction is repulsive [3]. 
Of course, lattice - gas models can give only a rough parody on real systems 
with more complicated structures, although these models often describe well 
chemical processes [32]. Low dimensionality (less than three) also makes it 
more difficult the appearance of thermal structural fluctuations. For 
instance, the latter, as has been rigorously shown [29,33 - 35], do not arise 
in the two - dimensional Ising - type models, though can exist in 
three - dimensional ones. Nevertheless, a frozen metastable coexistence of 
two different structures seems to be always possible. The condition for the 
appearance of thermal structural fluctuations changes to be more favorable in 
the vicinity of a structural phase transition where at least one of, or even 
both, coexisting structures become unstable and soft.

\section{Crystalline Structures}

Consider the case when two coexisting structures are both crystalline being 
characterized by different lattice vectors. Lattice sites of each structure 
are given by a vector set $\;\{\stackrel{\rightarrow}{a}_{i\nu}|\; i=1,2,
\ldots ,N \}\;$. As usual [36], we can expand 
$\;\Phi (\stackrel{\rightarrow}{R}_{ij\nu} )\;$ in Hamiltonian (15) in powers 
of the deviations
\begin{equation}
\stackrel{\rightarrow}{u}_{i\nu} \equiv \stackrel{\rightarrow}{R}_{i\nu} - 
\stackrel{\rightarrow}{a}_{i\nu} 
\end{equation}
from the lattice sites, limiting ourselves by the second order of this 
expansion. We are interested here in qualitive understanding of the behavior 
of a heterostructural system, therefore we will not discuss such questions 
that do not change principally this behavior although can be important in 
quantitative calculations for particular substances. These questions include 
the account of interparticle correlations [17,37 - 40] of anharmonicities 
[41,42], and of vacancies [43]. Instead, we can think of 
$\;\Phi (\stackrel{\rightarrow}{R}_{ij\nu}) \; $ as of an effective 
potential adjust to take into account these effects, at least partially.

After expanding Hamiltonian (15) in powers of deviations (21), we get
\begin{equation}
H_\nu = E_\nu^{st} + H_\nu^{ph} ,
\end{equation}
where the first term
\begin{equation}
E_\nu^{st} = Nw_\nu^2U_\nu
\end{equation}
is a static potential energy with
\begin{equation}
U_\nu \equiv\frac{1}{2N}\sum_{i\neq j}^{N}\Phi (
\stackrel{\rightarrow}{a}_{ij\nu} ) \qquad 
( \stackrel{\rightarrow}{a}_{ij\nu} \equiv \stackrel{\rightarrow}{a}_{i\nu} - 
\stackrel{\rightarrow}{a}_{j\nu} ) ,
\end{equation}
and the second  term
\begin{equation}
H_\nu^{ph} = w_\nu\sum_{i=1}^{N}\frac{p_{i\nu}^2}{2m} + 
\frac{w_\nu^2}{2}\sum_{i,j}^{N} 
D_{ij\nu}^{\alpha\beta} u_{i\nu}^\alpha u_{j\nu}^\beta 
\end{equation}
is the phonon Hamiltonian with the dynamical matrix
$$ D_{ij\nu}^{\alpha\beta} = \frac{\partial^2\Phi (
\stackrel{\rightarrow}{a}_{ij\nu} )}{\partial a_{i\nu}^\alpha
\partial a_{j\nu}^\beta} \qquad (i \neq j) $$
$$  D_{ii\nu}^{\alpha\beta} = \sum_{j(\neq i)}^{N}
\frac{\partial^2\Phi (\stackrel{\rightarrow}{a}_{ij\nu} )}
{\partial (a_{i\nu}^\alpha )^2} . $$
The mean potential energy (18) becomes
\begin{equation}
\Phi_\nu = U_\nu + \frac{1}{2N} \sum_{i,j}^{N} 
\sum_{\alpha ,\beta}^{3} D_{ij\nu}^{\alpha\beta} 
< u_{i\nu}^\alpha u_{j\nu}^\beta > .
\end{equation}
The Hamiltonian (25) can be diagonalized in a standard way by introducing the 
transformations
$$ \stackrel{\rightarrow}{u}_{i\nu} = \sum_{k,s} 
\frac{\stackrel{\rightarrow}{e}_{ks\nu}}{(2mN\omega_{ks\nu} )^{1/2}} 
\left ( b_{ks\nu} + b_{-ks\nu}^{+} \right ) e^{i\stackrel{\rightarrow}{k}
\stackrel{\rightarrow}{a}_{i\nu}} , $$
\begin{equation}
\stackrel{\rightarrow}{p}_{i\nu} = -i\sum_{k,s} 
\left (\frac{m\omega_{ks\nu}}{2N} \right )^{1/2} 
\stackrel{\rightarrow}{e}_{ks\nu}
\left ( b_{ks\nu} - b_{-ks\nu}^{+} \right ) 
e^{i\stackrel{\rightarrow}{k}\stackrel{\rightarrow}{a}_{i\nu}} , 
\end{equation}
using the orthogonality condition
\begin{equation}
\frac{1}{N} \sum_{i=1}^{N} \stackrel{\rightarrow}{e}_{ks\nu} 
\stackrel{\rightarrow}{e}_{k's'\nu} e^{i(\stackrel{\rightarrow}{k} - 
\stackrel{\rightarrow}{k}')\stackrel{\rightarrow}{a}_{i\nu}} = \delta_{kk'}
\delta_{ss'} , 
\end{equation}
and defining the frequencies $ \; \omega_{ks\nu} \; $ by the eigenvalue 
problem
\begin{equation}
\frac{w_\nu}{m} \sum_{j=1}^{N} \sum_{\beta =1}^3 D_{ij\nu}^{\alpha\beta} 
e^{i\stackrel{\rightarrow}{k}\stackrel{\rightarrow}{a}_{ij}} 
e_{ks\nu}^\beta = \omega_{ks\nu}^2 e_{ks\nu}^\alpha ,
\end{equation}
whose eigenfunctions are the polarization vectors 
$\;\stackrel{\rightarrow}{e}_{ks\nu} \; $. As a result, we have the phonon 
Hamiltonian
\begin{equation}
H_\nu^{ph} =w_\nu \sum_{k,s} \omega_{ks\nu} \left ( b_{ks\nu}^+b_{ks\nu} +
\frac{1}{2} \right ) .
\end{equation}

Although the way of obtaining (30) is standard, following it we have to be 
very cautious paying much attention to the nontrivial dependence of the 
Hamiltonian on the structure probability factors $ \; w_\nu \; $. Because of 
this, the phonon frequencies defined by (29) become dependent on $\; w_\nu\;$ 
as well. The structure probability $ \; w_\nu \; $ enters also in all main 
averages such as the phonon distribution
\begin{equation}
n_{ks\nu} \equiv < b_{ks\nu}^+ b_{ks\nu} > = \left [ 
\exp \left ( \frac{w_\nu\omega_{ks\nu}}{T} \right ) - 1 \right ]^{-1} ,
\end{equation}
the momentum squared
\begin{equation}
< \stackrel{\rightarrow}{p}_{i\nu}^2 > = \frac{m}{2N} \sum_{k,s} 
\omega_{ks\nu} \coth \frac{w_\nu\omega_{ks\nu}}{2T} ,
\end{equation}
and the correlation function
\begin{equation}
< u_{i\nu}^\alpha u_{j\nu}^\beta > = \frac{\delta_{ij}}{2N} \sum_{k,s} 
\frac{e_{ks\nu}^\alpha e_{ks\nu}^\beta}{m\omega_{ks\nu}} \coth 
\frac{w_\nu\omega_{ks\nu}}{2T} . 
\end{equation}
In its turn, Eq.(16) for the structure probability involves the mean kinetic energies (17),
\begin{equation}
K_\nu = \frac{1}{4N} \sum_{k,s} \omega_{ks\nu} \coth 
\frac{w_\nu\omega_{ks\nu}}{2T} ,
\end{equation}
and the mean potential energies (18),
$$ \Phi_\nu = U_\nu + D_\nu , $$
\begin{equation}
D_\nu = \frac{1}{4N} \sum_{j=1}^{N} \sum_{k,s} \sum_{\alpha ,\beta}^3 
\frac{e_{ks\nu}^\alpha e_{ks\nu}^\beta}{m\omega_{ks\nu}} 
D_{ij\nu}^{\alpha\beta} \coth \frac{w_\nu \omega_{ks\nu}}{2T} .
\end{equation}

The internal energy of the heterostructural system is
$$ E \equiv < \stackrel{-}{H} > = E_1 + E_2 , $$
\begin{equation}
E_\nu \equiv < H_\nu > = E_\nu^{st} + E_\nu^{ph} ,
\end{equation}
where the static energy is given by (23) and the phonon energy is 
\begin{equation}
E_\nu^{ph} \equiv < H_\nu^{ph} > = \frac{w_\nu}{2} \sum_{k,s} 
\omega_{ks\nu} \coth \frac{w_\nu\omega_{ks\nu}}{2T} .
\end{equation}
The latter, using the relations
\begin{equation}
K_\nu =w_\nu D_\nu , \qquad w_\nu\Phi_\nu = w_\nu U_\nu + K_\nu , 
\end{equation}
can be written as
\begin{equation}
E_\nu^{ph} = N ( w_\nu K_\nu + w_\nu^2 D_\nu ) = 2Nw_\nu K_\nu .
\end{equation}
Thus, we see that the internal energy (36) depends on temperature directly 
and also through the structure probabilities $ \; w_\nu \; $. Consequently, 
the specific heat of a heterophase system,
$$ C_\nu \equiv \frac{\partial E}{\partial T} = 
\left ( \frac{\partial E}{\partial T} \right )_w + \left ( 
\frac{\partial E}{\partial w} \right )_T , $$
contains an additional term, as compared to the specific heat of a pure 
single - structure system. This excessive term makes it possible to explain 
the so - called specific - heat anomalies observed in heterophase systems 
[24,44].

The free energy (7) takes the form
$$ f = f_1 + f_2 , $$
\begin{equation}
f_\nu = \frac{1}{N} E_\nu^{st} + \frac{T}{N} \sum_{k,s} \ln \left ( 2 \sinh 
\frac{w_\nu\omega_{ks\nu}}{2T} \right ) ,
\end{equation}
which demonstrates the nonlinear dependence on the structure probabilities 
$ \; w_\nu \; $.

\section{Structure Probabilities} 

The equation defining the structure probabilities can be written either as 
Eq.(16) with substituted there mean kinetic energies (34) and potential 
energies (35) or can be obtained by the direct minimization of (40) with 
respect to $ \; w \equiv w_1 \; $ and taking into account (6). Both ways, 
as can be checked, yield the same answer. To analyse this equation, we have 
first of all to remember that phonon frequencies, given by the eigenvalue 
problem (29), depend on structure probabilities. To make this dependence 
explicit, we introduce the notation
\begin{equation}
\omega_{ks\nu} \equiv w_\nu^{1/2}\varepsilon_{ks\nu} ,
\end{equation}
$$ \varepsilon_{ks\nu}^2 \equiv \frac{1}{m} 
\sum_{j=1}^{N} \sum_{\alpha ,\beta}^3 D_{ij\nu}^{\alpha\beta} 
e_{ks\nu}^\alpha e_{ks\nu}^\beta 
e^{i\stackrel{\rightarrow}{k} \stackrel{\rightarrow}{a}_{ij\nu}} , $$
in which $ \; \varepsilon_{ks\nu} \; $ does not contain $ \; w_\nu \; $. 
Emphasizing the dependence on $ \; w_\nu \; $ explicitly, we have for the 
kinetic energy (34)
\begin{equation}
K_\nu = \frac{w_\nu^{1/2}}{4N} \sum_{k,s} \varepsilon_{ks\nu}\coth 
\frac{w_\nu^{3/2}\varepsilon_{ks\nu}}{2T} .
\end{equation}
For the internal energy (36) we get
$$ E = E_1 + E_2 , \qquad E_\nu = Nw_\nu^2 U_\nu + 2Nw_\nu K_\nu . $$
The free energy (40) becomes 
$$ f = f_1 + f_2 , $$
\begin{equation}
f_\nu = w_\nu^2 U_\nu + \frac{T}{N} \sum_{k,s} \ln \left ( 2\sinh 
\frac{w_\nu^{3/2}\varepsilon_{ks\nu}}{2T} \right ) .
\end{equation}
Minimizing (43) with respect to $ \; w \; $, with the use of notation (8) 
and relations
$$ \frac{\partial f}{\partial w} = \frac{\partial f}{\partial w_1} - 
\frac{\partial f}{\partial w_2}, \qquad \frac{\partial f}{\partial w_\nu} = 
2w_\nu U_\nu + 3K_\nu , $$
we obtain
\begin{equation}
w = \frac{2U_2 +3(K_2 - K_1)}{2(U_1 + U_2)} .
\end{equation}
From the inequalities $ \; 0 \leq w \leq 1 \; $, assuming that
\begin{equation}
U_1 + U_2 < 0 
\end{equation}
we have a necessary condition
\begin{equation}
U_2 \leq \frac{3}{2} \left ( K_1 - K_2 \right ) \leq - U_1
\end{equation}
for $ \; w \; $ to be considered as a probability.

To further simplify the analyses, let us resort to the Debye approximation. 
For this, we pass to the thermodynamic limit by using the change
$$ \frac{1}{N} \sum_{k,s} \rightarrow \frac{1}{\rho} \sum_{s=1}^3 
\int_{\bf D} \frac{d\stackrel{\rightarrow}{k}}{(2\pi)^3} , $$ where
$$ {\bf D} \equiv \{ \stackrel{\rightarrow}{k} | \; k \equiv 
|\stackrel{\rightarrow}{k} | \leq k_D \} $$
is the Debye sphere, and the Debye momentum $ \; k_D \; $ being defined by 
the normalization $ \; \frac{1}{N}\sum_{k,s} 1 = 1 \; $ giving 
$$ k_D = (6\pi^2\rho )^{1/3}, \qquad \rho \equiv \frac{N}{V} . $$
The phonon spectrum in the Debye approximation acquires the linear form, for 
which one has to make the substitution 
$$ \varepsilon_{ks\nu} \rightarrow \varepsilon_{k\nu} = c_\nu k , $$
$$ c_\nu^2 \equiv \lim_{k \rightarrow 0} \frac{1}{3} \sum_{s=1}^3 \left ( 
\frac{\varepsilon_{ks\nu}}{k} \right )^2 . $$
The quantity $ \; c_\nu \; $ plays the role of the sound velocity in a pure 
$ \; \nu \; $- structure. From (41), using the orthogonality property
$$ \sum_{s=1}^3 e_{ks\nu}^\alpha e_{ks\nu}^\beta = \delta_{\alpha\beta} , $$
one gets
\begin{equation}
c_\nu^2 = -\lim_{k \rightarrow 0} \sum_{j=1}^{N} \sum_{\alpha =1}^{3} 
D_{ij\nu}^{\alpha\alpha} 
\frac{(\stackrel{\rightarrow}{k}\stackrel{\rightarrow}{a}_{ij\nu})^2}{6mk^2} .
\end{equation}
In this way, for the kinetic energy (42) we have
\begin{equation}
K_\nu = \frac{36T^4}{\Theta_\nu^3w_\nu} \int_{0}^{\Theta_\nu /2T} x^3 
\coth x dx
\end{equation}
and for the free energy $ \; f_\nu \; $ we get
\begin{equation}
f_\nu = w_\nu^2 U_\nu + \frac{72T^4}{\Theta_\nu^3} \int_{0}^{\Theta_\nu /2T} 
x\ln (2\sinh x )dx ,
\end{equation}
where we have introduced the notation
\begin{equation}
\Theta_\nu \equiv w_\nu^{3/2}T_{\nu D} \qquad (T_{\nu D} \equiv c_\nu k_D ) .
\end{equation}
Here, $ \; T_{\nu D} \; $ is the Debye temperature of a pure $ \; \nu \; $- 
structure, while $ \; \Theta_\nu \; $ can be called an effective Debye 
temperature of a structure inside a mixed heterostructural system.

Formulas (48) - (50) show that the low and high temperature limits for a 
heterostructural system are to be defined not with respect to $\;T_{\nu D}\;$ 
but with respect to the effective temperature $ \; \Theta_\nu \; $ given by 
(50). The latter is renormalized by the factor $ \; w_\nu^{3/2} \; $ itself 
depending on temperature. To analyse the behavior of the structural 
probability (44) we need to write accurately the corresponding temperature 
limits for the kinetic energy (48).

In the case, when
$$ T \ll \frac{\Theta_\nu}{2\pi} = \frac{w_\nu^{3/2}}{2\pi}T_{\nu D} , $$
we can use the integrals
$$ \int_{0}^{\infty}\frac{x^{2n-1}}{e^x -1}dx = (-1)^{n-1} 
\frac{(2\pi)^{2n}}{4n}B_{2n} , $$
in which $ \; B_n \; $ are the Bernoulli numbers,
$$ B_0 = 1, \; B_1 =-\frac{1}{2} , \; B_2 = \frac{1}{6} , \; B_3 = 0 , 
\; B_4 = -\frac{1}{30}, \ldots $$
In particular,
$$ \int_{0}^{\infty} \frac{x^3dx}{e^x - 1} = \frac{\pi^4}{15} , \qquad 
\int_{0}^{\infty} \frac{xdx}{e^x -1} = \frac{\pi^2}{6} . $$
This yields for the kinetic energy (48)
\begin{equation}
K_\nu \simeq \frac{9}{16} T_{\nu D} w_\nu^{1/2} + 
\frac{3\pi^4T^4}{10T_{\nu D}^3w_\nu^{1/2}} .
\end{equation}
 
In the opposite limit, when
$$ T \gg \frac{\Theta_\nu}{2\pi} = \frac{w_\nu^{3/2}}{2\pi}T_{\nu D} , $$
using the expansion
$$ \coth x \simeq \frac{1}{x} + \frac{x}{3} - \frac{x^3}{45} 
\qquad ( x < \pi ) , $$
we find
\begin{equation}
K_\nu \simeq \frac{3T}{2w_\nu} + \frac{3T_{\nu D}^2}{40T} w_\nu^2 .
\end{equation}

To make the following expressions less cumbersome, it is convenient to 
introduce the dimensionless static energies
$$ u_1 \equiv \frac{U_1}{T_{1D}} , \qquad u_2 \equiv \frac{U_2}{T_{1D}} , $$
\begin{equation}
u \equiv - \frac{U_1 + U_2}{T_{1D}} = - (u_1 + u_2 ).
\end{equation}
In the case of (45), the latter value is positive, $ \; u > 0 \; $. Also, we 
shall use the notation
\begin{equation}
t \equiv \frac{T}{T_{1D}} , \qquad \tau \equiv \frac{T_{2D}}{T_{1D}} .
\end{equation}
For definiteness, we assume that the structure corresponding to 
$ \; \nu = 1 \; $ has a higher Debye temperature, that is, 
$ \; T_{1D} > T_{2D} \; $. Hence, the parameter $ \; \tau \; $ from (54) 
lies in the region $ \; 0 < \tau < 1 \; $.

Now, let us understand the behavior of the structural probability (44) in the 
vicinity of the temperature $ \; T_s \; $ of a structural phase transition. 
This can be a phase transition either between two different crystalline 
structures or between a regular crystalline structure and an irregular glassy 
structure. The latter consideration is possible owing to many similarities 
between crystalline and glassy states [45,46] and because the Debye 
approximation is applicable to both of them. The qualitative behavior of the 
structural probabilities near the structural transition temperature $\;T_s\;$ 
is mainly influenced by the relation between $ \; T_s \; $ and the effective 
temperatures (50). It is possible to distinguish three cases that can be 
conditionally called the low - temperature, mid - temperature and high - 
temperature cases.

Begin with the low - temperature situation when the structural transition 
temperature satisfies the inequality
\begin{equation}
T_s < \frac{w_\nu^{3/2}}{2\pi} T_{\nu D} \qquad ( \nu = 1,2 ) .
\end{equation}
Then, for the kinetic energies of both structures we can use the approximation
(51) which is to be substituted into (44). To simplify the resulting 
expression, we notice that as $ \; w_\nu \leq 1 \; $ and 
$ \; T \approx T_s \; $, hence the variable $ \; t \; $ defined in (54) can 
be considered, according to (55), as  small parameter, since 
$ \; t < 1/2\pi = 0.159 \; $. This yields
\begin{equation}
w \simeq \alpha_0 - \alpha_4t^4 ,
\end{equation}
where $ \; \alpha_0 \; $ is a solution of the equation
$$ u_1\alpha_0 + \frac{27}{32}\sqrt{\alpha_0} = u_2(1 -\alpha_0) + 
\frac{27}{32}\tau\sqrt{1 -\alpha_0} , $$
and 
$$ \alpha_4 = 
\frac{144\pi^4(\sqrt{\alpha_0} -\tau^3\sqrt{1 -\alpha_0})}{5\tau^3 [ 64u
\sqrt{\alpha_0(1-\alpha_0)} - 27 (\sqrt{1-\alpha_0} + 
\tau \sqrt{\alpha_0})]} . $$
Remind that $ \; w \equiv w_1 \; $ corresponds to a more rigid structure 
for which $ \; T_{1D} > T_{2D} \; $. Eq.(56) shows that the probability of 
the more rigid structure quickly decreases as temperature increases in the 
vicinity of the structural - transition temperature $ \; T_s \; $.

Consider now the case when the transition temperature $ \; T_s \; $ is, in 
some sense, intermediate satisfying the condition
\begin{equation}
\frac{w_2^{3/2}}{2\pi}T_{2D} < T_s < \frac{w_1^{3/2}}{2\pi}T_{1D} .
\end{equation}
Then, the kinetic energy of the more rigid structure can be approximated by 
Eq.(51) while that of the more soft is to be approximated by (52), which gives
$$ K_1 \simeq \frac{9}{16}T_{1D}\sqrt{w} + 
\frac{3\pi^4T^4}{10T_{1D}^3\sqrt{w}} , $$
$$ K_2 \simeq \frac{3T}{2(1 - w)} + \frac{3T_{2D}^2}{40T} (1 - w)^2. $$
Substituting this into (44), we find
\begin{equation}
w \simeq 1 - \alpha_1t ,
\end{equation}
where
$$ \alpha_1 = \frac{72}{32u_1 + 27} . $$
Now again the probability of the more rigid structure decreases with 
increasing temperature, although not so quickly as in (56). The solution 
(58) exists only if $ \; \alpha_1 > 0 \; $. If $ \; \alpha_1 \leq 0 \; $, 
we have to put $ \; w = 1 \; $, which means that there are no structural 
fluctuations.

Finally, pass to the high - temperature case, when
\begin{equation}
T_s > \frac{w_\nu^{3/2}}{2\pi}T_{\nu D} \qquad (\nu = 1,2) .
\end{equation}
For the kinetic energies of both structures we can use the approximation (52).
Then, (44) yields
\begin{equation}
w \simeq \frac{1}{2} + \frac{\beta_1}{t}
\end{equation}
for $ \; t \gg 1 \; $ and
$$ \beta_1 = \frac{1}{36} \left ( u_1 - u_2 \right ) . $$
In the vicinity of the structural phase transition the structure probabilities
are close to each other, $ w \approx 1/2 \; $. The latter equation becomes 
asymptotically exact if $ \; u_1 \rightarrow u_2 \; $. This is analogous to 
the case (55) for which (56) also gives $ \; w \approx 1/2 \; $ if the 
properties of both coexisting structures are similar to each other, that is 
if $ \; \tau \approx 1 \; $ and $ \; u_1 \approx u_2 \; $.

The above analysis shows that the appearance of thermal structural 
fluctuations near the point of a structural phase transition is facilitated 
when both coexisting structures have close characteristics.

\section{M\"ossbauer Factor}

The occurrence of structural fluctuations around the point of a structural 
transition can lead to the emergence of various anomalies of observable 
quantities [22], such as the strong enhancement of specific heat and of 
diffusion coefficient [22,24,47]. A detailed analysis of experimental data 
[48 - 51] confirms that, probably, the most common feature of structural 
transitions is an anomalous sagging of the M\"ossbauer factor near the 
transition point. Such a sagging, as has been proved [52,53], cannot be 
explained by the existence of a soft phonon mode, which can lead solely to a 
fracture of the M\"ossbauer factor but by no means to a sagging. However, the 
general softening of a crystal due to the arising structural fluctuations can 
provoke these saggings, as we demonstrate below.

The M\"ossbauer factor of a heterostructural system consisting of two 
thermally coexisting structures  is written in the form
\begin{equation}
f_M = | w_1\varphi_1 + w_2\varphi_2 |^2 = f_M(T,w) ,
\end{equation} 
where $ \; w = w(T) \; $ is the structure probability given by (44), and
$$ \varphi_\nu \equiv e^{-W_\nu} =\varphi_\nu (T,w_\nu ) , $$
\begin{equation}
W_\nu = \frac{q^2}{2}r_\nu^2 = E_qmr_\nu^2 \qquad (E_q \equiv 
\frac{q^2}{2m} ) ,
\end{equation}
here $ \; E_q \; $ is the recoil energy; $ \; q \; $, the absolute value of a 
gamma - quantum momenta; and
\begin{equation}
r_\nu^2 =\frac{1}{3} \sum_{\alpha=1}^3 <u_{i\nu}^\alpha u_{i\nu}^\alpha >
\end{equation}
is the mean - square oscillation amplitude of a particle in a $ \; \nu \; $- 
structure. The correlation function (33) in the Debye approximation is
\begin{equation}
< u_{i\nu}^\alpha u_{i\nu}^\beta > = \delta_{\alpha\beta} 
\frac{6T^2w_\nu}{m\Theta_\nu^3} \int_{0}^{\Theta_\nu /2T} x\coth x dx .
\end{equation}
Whence, the mean - square amplitude (63) becomes
\begin{equation}
r_\nu^2 = \frac{6T^2w_\nu}{m\Theta_\nu^3} \int_{0}^{\Theta_\nu /2T} x
\coth x dx .
\end{equation}
The low - and high - temperature asymptotes for (65) are 
$$ r_\nu^2 \simeq \frac{w_\nu}{m\Theta_\nu} \left ( \frac{3}{4} + 
\frac{\pi^2T^2}{2\Theta_\nu^2} \right ) \qquad 
( T \ll \frac{\Theta_\nu}{2\pi} ) , $$
\begin{equation}
r_\nu^2 \simeq \frac{w_\nu}{m\Theta_\nu} \left ( \frac{3T}{\Theta_\nu} + 
\frac{\Theta_\nu}{12T} \right ) \qquad 
( T \gg \frac{\Theta_\nu}{2\pi} )  .
\end{equation}
Using (66), and remembering formula (50), for the function 
$ \; \varphi_\nu \; $ defined in (62) we have
$$ \varphi_\nu \simeq \exp \left ( - 
\frac{3E_q}{4T_{\nu D}w_\nu^{1/2}} \right )
\qquad \left ( T < \frac{\Theta_\nu}{2\pi} \right ) , $$
\begin{equation}
\varphi_\nu \simeq \exp \left ( - \frac{3TE_q}{T_{\nu D}^2w_\nu^2} \right )
\qquad \left ( T > \frac{\Theta_\nu}{2\pi} \right ) .
\end{equation}

In order to elucidate the influence of structural fluctuations appearing near 
the temperature $ \; T_s \; $ of a structural transition, we will compare the 
M\"ossbauer factor (61) of a heterostructural system with the M\"ossbauer 
factor
\begin{equation}
f_M(T,0) \equiv \exp \left \{ -12E_q \frac{T^2}{T_{2D}^3} 
\int_{0}^{T_{2D} /2T} x\coth x dx \right \} 
\end{equation}
of a pure high - temperature structure corresponding to $ \; \nu = 2 \; $. 
The asymptotic values of the reference factor (68) are
$$ f_M(T,0) \simeq \exp \left ( -\frac{3E_q}{2T_{2D}} \right ) \qquad 
\left ( T < \frac{T_{2D}}{2\pi} \right ) , $$
\begin{equation}
f_M(T,0) \simeq \exp \left ( -\frac{6TE_q}{2T_{2D}^2} \right ) \qquad 
\left ( T > \frac{T_{2D}}{2\pi} \right ) .
\end{equation}
The change of the M\"ossbauer factor influenced by the presense of structural 
fluctuations is convenient to characterize by the relative deviation 
\begin{equation}
\delta f_M(T,w) \equiv \frac{f_M(T,w)}{f_M(T,0)} - 1.
\end{equation}

Consider first the case when the temperature of a structural transition is 
such that
\begin{equation}
T_s < \frac{\Theta_\nu}{2\pi} \qquad ( \nu = 1,2 ) .
\end{equation}
For $ \; T \approx T_s \; $ we have
$$ \varphi_1(T,w) \simeq [f_M(T,0)]^{\tau /2\sqrt{w}} , $$
$$ \varphi_2(T,w) \simeq [f_M(T,0)]^{1/2\sqrt{1-w}} . $$
Therefore, for the relative change (70) we get
\begin{equation}
\delta f_M(T_s,w) \simeq \frac{1}{f_0} \left | wf_0^{\tau /2\sqrt{w}} + 
(1 - w)f_0^{1/2\sqrt{1-w}} \right |^2 - 1 ,
\end{equation}
where
$$ f_0 \equiv f_M(T_s,0) , \qquad w = w(T_s) . $$
If we assume that $ \; w \approx 1/2 \; $, then (72) transforms to
$$ \delta f_M \left ( T_s,\frac{1}{2} \right ) \simeq \frac{1}{4f_0} 
\left ( f_0^{\tau /\sqrt{2}} + f_0^{1/\sqrt{2}} \right )^2 - 1. $$
When the two coexisting structures are drastically different, so that 
$ \; T_{1D} \gg T_{2D} \; $, that is $ \; \tau \ll 1 \; $, then
$$ \delta f_M \left ( T_s, \frac{1}{2} \right ) \simeq \frac{1}{4f_0} 
\left ( 1 + f_0^{0.707} \right )^2 - 1 \qquad ( \tau \ll 1 ) , $$
and when they are similar, so that $ \; \tau \approx 1 \; $, then
$$ \delta f_M \left ( T_s , \frac{1}{2} \right ) \simeq f_0^{0.414} - 1 
\qquad ( \tau \approx 1 ) . $$
To estimate these quantities, let us take the reference M\"ossbauer factor 
$ \; f_0 \; $ from the region
\begin{equation}
0.7 \leq f_0 \leq 0.9 .
\end{equation}
Then, we obtain
$$ 0.033 \leq \delta f_M \left ( T_s,\frac{1}{2} \right ) \leq 0.128 \qquad
(\tau \ll 1 ) , $$
\begin{equation}
- 0.137 \leq \delta f_M \left ( T_s,\frac{1}{2} \right ) \leq - 0.043 \qquad
(\tau \approx 1 ) .
\end{equation}
As we see, a sagging at $ \; T_s \; $ can be directed upward as well as 
downward, depending on the parameter $ \; \tau \equiv T_{2D}/T_{1D} \; $ 
characterizing the difference between the structures. When 
$ \; \tau \approx 0.5 \; $, there is no sagging at all.

Turn now to the case when the temperature of a structural transition is in 
the interval
\begin{equation}
\frac{\Theta_2}{2\pi} < T_s < \frac{\Theta_1}{2\pi}.
\end{equation}
Hence, at $ \; T \approx T_s \; $ one gets
$$ \varphi_1(T,w) \simeq [ f_M(T,0)]^{\tau^2 /8t\sqrt{w}} , $$
$$ \varphi_2(T,1-w) \simeq [f_M(T,0)]^{1/2(1-w)^2} . $$
For the relative change (70) we find 
\begin{equation}
\delta f_M(T_s,w) =\frac{1}{f_0} \left | wf_0^{w^2/8t_s\sqrt{w}} +
(1 - w) f_0^{1/2(1-w)^2} \right |^2 - 1
\end{equation}
When $ \; T_s \; $ is in the middle of the interval (75), we can use the 
approximation
$$ t_s \equiv \frac{T_s}{T_{1D}} \approx 
\frac{1}{2} \left ( 1 + \tau \right ) . $$
Taking $ \; w \approx 1/2 \; $, we reduce (76) to
$$ \delta f_M \left ( T_s , \frac{1}{2} \right ) \simeq \frac{1}{4f_0} 
\left ( f_0^{\tau^2/2\sqrt{2}(1+\tau)} + f_0^2 \right )^2 - 1 . $$
Considering again two limiting situations of very different and similar 
structures, we have
$$ \delta f_M \left ( T_s, \frac{1}{2} \right ) \simeq \frac{1}{4f_0} 
\left ( 1 + f_0^2 \right )^2 - 1 \qquad (\tau \ll 1 ) , $$
$$ \delta f_M \left ( T_s, \frac{1}{2} \right ) \simeq \frac{1}{4f_0} \left (
f_0^{0.177} + f_0^2 \right )^2 - 1 \qquad (\tau \approx 1 ) . $$
Taking the values of $ \; f_0 \; $ from (73), we come to the result
$$ -0.207 \leq \delta f_M \left ( T_s, \frac{1}{2} \right ) \leq -0.090 
\qquad (\tau \ll 1 ) , $$
\begin{equation}
-0.271 \leq \delta f_M \left ( T_s, \frac{1}{2} \right ) \leq -0.108 \qquad
(\tau \approx 1 ) .
\end{equation}
Here the sagging are always directed downward.

If the structural transition occurs at high temperature, when
\begin{equation}
T_s > \frac{\Theta_\nu}{2\pi} \qquad ( \nu = 1,2 ) ,
\end{equation}
then
$$ \varphi_1(T,w) \simeq [f_M(T,0)]^{\tau^2/2w^2} , $$
$$ \varphi_2(T,w) \simeq [f_M(T,0)]^{1/2(1-w)^2} $$
at $ \; T \approx T_s \; $. The relative change (70) becomes
\begin{equation}
\delta f_M(T_s,w) \simeq \frac{1}{f_0} \left | wf_0^{\tau^2/2w^2} + 
(1 - w ) f_0^{1/2(1 - w)^2} \right |^2 - 1 .
\end{equation} 
For $ \; w \approx 1/2 \; $ Eq.(79) gives
$$ \delta f_M \left ( T_s, \frac{1}{2} \right ) \simeq \frac{1}{4f_0} \left (
f_0^{2\tau^2} + f_0^2 \right )^2 - 1 , $$
from where
$$ \delta f_M \left ( T_s, \frac{1}{2} \right ) \simeq \frac{1}{4f_0} \left (
1 + f_0^2 \right )^2 - 1 \qquad (\tau \ll 1 ) , $$
$$  \delta f_M \left ( T_s, \frac{1}{2} \right ) \simeq f_0^3 - 1 \qquad
(\tau \approx 1 ) . $$
Thence, invoking (73), we obtain
$$ -0.207 \leq \delta f_M \left ( T_s, \frac{1}{2} \right ) \leq -0.090
\qquad (\tau \ll 1 ) , $$
\begin{equation}
-0.657 \leq \delta f_M \left ( T_s, \frac{1}{2} \right ) \leq -0.271
\qquad (\tau \approx 1 ) .
\end{equation}
All saggings are again directed downward.

Thus, we see that the appearance of structural fluctuations near the 
temperature of a structural phase transition can yield a noticeable sagging 
of the M\"ossbauer factor as a function of temperature. In the majority of 
cases this sagging is directed downwards. The sagging can be easily observed 
if $ \; w \approx 1/2 \; $, and immediately disappears when any of the 
structure probabilities tends to zero. For example, suppose that the 
probability of the structure fluctuations of a high - temperature more soft 
phase inside a low - temperature more rigid structure is very small, so that 
$ \; w_2 \equiv x \ll 1 \; $. Expanding (61) in powers of $ \; x \; $, we have
$$ f_M (T, 1-x) \simeq f_M (T,1) \left [ 1 - \left ( 2 + \frac{3E_q}{2T_{1D}} 
\right ) x \right ] . $$
In the standard M\"ossbauer experiments the recoil energy is negligibly small 
as compared to the Debye temperature, $ \; E_q \ll T_{1D} \; $, usually 
$ \; E_q / T_{1D} \sim 10^{-7} \; $. Therefore, the change of the 
M\"ossbauer factor
$$ f_M (T, 1 - x ) \simeq f_M (T,1) ( 1 - 2x ) \qquad ( x \ll 1 )  $$
is also quite small and disappears as soon as $ \; x \rightarrow 0 \; $.

To illustrate that the values of saggings at $ \; T_s \; $ are in agreement 
with experiment, let us consider the structural transition between the low - 
temperature cubic phase and the high - temperature rhombic phase in the 
compounds $ \; Sn_{1-x}Ge_xTe \; $. The characteristic temperatures of the 
latter are $ \; T_s \approx 190 K \; $ and $ \; T_{1D} \approx T_{2D} \approx
150 K \; $. M\"ossbauer investigation [54] display the existence at 
$ \; T_s \; $ of a pronounced sagging of the M\"ossbauer factor, 
$ \; \delta f_M^{exp} (T_s ) \approx - 0.4 \; $. The considered characteristic
temperatures correspond to inequality (78) and to $ \; \tau \approx 1 \; $. 
The measured M\"ossbauer sagging is in agreement with the second line of 
estimate (80).

\section{Sound Velocity}

The existence of structural fluctuations near a phase transition point can 
also lead to a distinct attenuation of the velocity of sound 
\begin{equation}
v_s = \lim_{k \rightarrow 0} \frac{1}{k} \sum_{\nu} 
\frac{\delta}{\delta n_{ks\nu}} < H_\nu^{ph} > .
\end{equation}
To trace out the dependence of the sound velocity (81) on the structure 
probabilities, let us introduce the notation
\begin{equation}
c_{s\nu} \equiv \lim_{k \rightarrow 0} \frac{\varepsilon_{ks\nu}}{k}
\end{equation}
for the sound velocity with polarization $ \; s \; $ inside a pure structure 
$ \; \nu \; $. According to (30) and (31), the sound velocity (81) in a 
heterostructural substance takes the form
\begin{equation}
v_s = w_1^{3/2} c_{s1} + w_2^{3/2} c_{s2} .
\end{equation}
Defining the average, with respect to polarizations, velocities of sound for 
a heterostructural system,
\begin{equation}
v \equiv \frac{1}{3} \sum_{s=1}^3 v_s ,
\end{equation}
and for a pure $ \; \nu \; $- structure,
$$ c_\nu \equiv \frac{1}{3} \sum_{s=1}^3 c_{s\nu} , $$
we get
\begin{equation}
v = w_1^{3/2}c_1 + w_2^{3/2}c_2 ,
\end{equation}
where each of $ \; c_\nu \; $ is given by (47).

The relative decrease of the sound velocity, due to arising structural 
fluctuations, can be characterized by
\begin{equation}
\delta v(w) \equiv \frac{v}{c_2} - 1 .
\end{equation}
In the Debye approximation we have
$$ \frac{c_2}{c_1} = \frac{T_{2D}}{T_{1D}} \equiv \tau . $$
Therefore, (86) gives
\begin{equation}
\delta v(w) = \frac{1}{\tau} w^{3/2} + ( 1 - w)^{3/2} - 1 .
\end{equation}
Assuming that at the temperature $ \; T_s \; $ of a structural transition 
one has $ \; w \approx 1/2 \; $ and $ \; \tau \approx 1 \; $, we obtain
$$ \delta v \left ( \frac{1}{2} \right ) \approx - 0.293 \qquad (T = T_s) . $$
This decrease of the sound velocity is completely due to the onset of 
mesoscopic structural fluctuations and can happen even at first - order phase 
transitions when there are no microscopic critical fluctuations [55] related 
to second - order phase transitions. For example, a similar decrease of the 
sound velocity occurres at freezing point of water [56].

As far as the average sound velocity (84) can be expressed through the 
derivative
$$ \frac{\partial P}{\partial \rho} = mv^2 $$
of pressure with respect to density, and the same derivative is involved 
into the definition of the isothermal compressibility
$$ \kappa_T \equiv - \left ( V\frac{\partial P}{\partial V} \right )^{-1} =
\left ( \rho \frac{\partial P}{\partial \rho} \right )^{-1} , $$
we can easily find the influence of structural fluctuations on the latter. 
Thus, introducing the isothermal compressibility of pure structures,
$$ \kappa_\nu \equiv \frac{1}{m\rho c_\nu^2} , $$
for a heterostructural system we find
\begin{equation}
\kappa_T = \left [ \left ( \frac{w_1^3}{\kappa_1} \right )^{1/2} + \left ( 
\frac{w_2^3}{\kappa_2} \right )^{1/2} \right ]^{-2} .
\end{equation} 
Defining the relative change of compressibility by
\begin{equation}
\delta \kappa_T (w) \equiv \frac{\kappa_T}{\kappa_2} - 1 ,
\end{equation} 
and using the previous notation, we come to
\begin{equation}
\delta \kappa_T(w) = \frac{\tau^2}{[w^{3/2} + \tau (1 - w )^{3/2}]^2} - 1 .
\end{equation} 
If at the point of a structural phase transition we have $\; w\approx 1/2\;$ 
and $ \; \tau \approx 1 \; $, then
$$ \delta \kappa_T \left ( \frac{1}{2} \right ) \approx 1 
\qquad ( T = T_s ) . $$
This means a strong increase of compressibility.

\section{Conclusion} 

The main aim of this paper has been to present a general approach for 
describing statistical properties of heterostructural systems. This approach 
can be used for treating heterogeneous system with frozen structural 
fluctuations induced e.g. by shock waves or irradiation. Then, the relative 
volumes occupied by each structure, or the structure probabilities, are 
additional thermodynamic variables which can be defined experimentally by 
different means, for example by nuclear gamma resonance [4,5]. Probably, the 
most promising application of this approach is to considering substances with 
thermal structural fluctuations. This is of great importance for describing 
systems with structural phase transitions. The appearance of structural 
fluctuations around the phase transition point leads to various 
pretransitional, or precursor, phenomena that are often manifested in 
pronounced anomalies of thermodynamic and dynamic characteristics. The 
liquid - solid phase transition can be regarded as a kind of such structural 
transitions [16,17]. There are other numerous examples of structural 
transitions accompanied by the occurrence of structural fluctuations whose 
existence can be observed by different experiments. Some of the examples have 
been discussed in this paper.

A number of other examples has been reviewed recently [22]. As is well known, 
structural fluctuations play a decisive role in high - temperature 
superconductors [23], in solids with martensitic transformations [57,58], 
and in crystals with perovskite structure whose structural transitions are 
accompanied by experimentally well observed pretransitional structural 
fluctuations [59,60]. A more detailed theoretical consideration of these 
particular substances is supposed to be done in separate publications, basing 
on the approach developed in the present paper.

\vspace{1cm}

{\bf Acknowledgements}

I am grateful to A.J.Coleman for the interest to my work and advises. 
Financial support by the Natural Sciences and Engineering Research Council of 
Canada is warmly appreciated.

\begin{center}
{\bf Figure Captions}
\end{center}

{\bf Fig.1}. The typical distribution of defect clusters inside a metal 
irradiated by fast neutrons.

\vspace{5mm}

{\bf Fig.2}. Pores and cracks in a metal irradiated by fast neutrons: darker 
regions correspond to the amorphised phase with disordered structure.

\end{document}